\documentclass{JAC2003}
\usepackage{graphicx}
\usepackage{booktabs}

\begin{document}
\title{DIHEDRAL GROUP AND REPETITIVE ACHROMATS WITH MIRROR SYMMETRIC
OR MIRROR ANTISYMMETRIC BASIC CELL
\vspace{-0.5cm}}
\author{V.Balandin\thanks{vladimir.balandin@desy.de}, 
R.Brinkmann, W.Decking, N.Golubeva \\
DESY, Hamburg, Germany}

\maketitle

\begin{abstract}
\vspace{-0.15cm}
Using the group-theoretical point of view we study in this paper second and 
third order repetitive achromats with a mirror symmetric or mirror antisymmetric 
basic cell and compare these achromats with repetitive achromats designed 
without internal cell symmetries taken into account.   
\end{abstract}

\vspace{-0.22cm}
\section{INTRODUCTION}

\vspace{-0.1cm}
As an achromat we understand a particle transport system whose linear 
transfer matrix is dispersion free (first-order achromat) and whose transfer 
map does not have nonlinearities of transverse motion up to a certain order 
$m$ ($m$-order achromat). 
The first practical solution for the second-order achromat was presented at 
the end of 1970s in the paper ~\cite{Brown}, where the theory of achromats 
based on repetitive symmetry was developed, and quickly becomes part of many 
accelerator designs. 
The way to construct the third and higher order repetitive achromats was pointed 
out about ten years later in ~\cite{Dragt} and was based on normal form theory.
The capabilities of the mirror (reversal) symmetry for the cancellation of 
nonlinear aberrations were explored in ~\cite{Wan_Berz} and, taking additionally 
into account the magnifying achromat ~\cite{Brown_Servranckx} and the four-cell 
staircase achromat mentioned, for example, in ~\cite{Rusthoi_Wadlinger}, the 
theory of achromats eventually came to the state when one has a set of 
different design recipes without a single theoretical framework and 
no answer to the question, if there is a magnet 
arrangement which will give better automatic cancellation of aberrations than 
those already known.

The way to create the unified theory of achromats was found in ~\cite{AchromIPAC10} 
by looking at the design of the symmetry based achromats  from the point of view of 
the theory of finite matrix groups. This approach not only explains why all previously known 
designs work, but also says that there is no better cancellation 
of aberrations than cancellation provided by the action of finite cyclic or dihedral 
group (at least as long as the linear transverse oscillations are uncoupled).
In this paper we compare the actions of cyclic and dihedral 
groups using as test bed second and third order repetitive achromats with a mirror 
symmetric or mirror antisymmetric basic cell.
Note that these achromats are new by itself and were not considered before
with the exception of the two-cell case, which was studied in ~\cite{Wan_Berz} 
under simplifying assumption that the cell matrix is dispersion free.

\vspace{-0.22cm}
\section{VARIABLES, MAPS AND ACHROMATS}

\vspace{-0.12cm}
We consider the beam dynamics in a magnetostatic system
which is symmetric about the horizontal midplane $y=0$ 
and use a complete set of symplectic variables
$\mbox{\boldmath $z$} = (x, p_x, y, p_y, \sigma, \varepsilon)^{\top}$
as particle coordinates. 
In this set the variables $\hat{\mbox{\boldmath $z$}} = (x, p_x, y, p_y)^{\top}$
describe the transverse particle motion and 
the variables $\sigma$ and $\varepsilon$ characterize
the longitudinal dynamics ~\cite{AchromIPAC10}.
We represent particle transport from one longitudinal location to another
by a symplectic map and assume that for arbitrary two longitudinal positions 
the point $\mbox{\boldmath $z$} = \mbox{\boldmath $0$}$ is the fixed
point and that the corresponding map can be Taylor 
expanded in its neighborhood.
We use that up to any predefined order $m$ the aberrations 
of a map ${\cal M}$ can be represented through a Lie factorization as 

\vspace{-0.2cm}
\noindent
\begin{eqnarray}
:{\cal M}: \,=_m\,
\exp(:{\cal F}_{3, m + 1}:) :M: ,
\label{IFB_3}
\end{eqnarray}

\vspace{-0.2cm}
\noindent
where 
${\cal F}_{3, m + 1} \stackrel{\mbox{\tiny def}}{=} {\cal F}_{3} + \ldots + {\cal F}_{m + 1}$,
each of the functions ${\cal F}_k$ is a homogeneous polynomial of degree $k$ in the variables $\mbox{\boldmath $z$}$ and the symbol $=_m$ denotes equality up to order $m$ (inclusive)
when maps on both sides of (\ref{IFB_3})
are applied to the phase space vector $\mbox{\boldmath $z$}$.

We also use that the map ${\cal M}$ of a magnetic system  which is 
symmetric about the horizontal midplane $\,y = 0\,$ satisfies

\vspace{-0.2cm}
\noindent   
\begin{eqnarray}
:{\cal M}:\,:T_M: \,=\,: T_M: \,:{\cal M}: ,
\label{TWO_C2_6}
\end{eqnarray}

\vspace{-0.2cm}
\noindent   
where $T_M = \mbox{diag}(1, 1, -1, -1, 1, 1)$
is the mid-plane symmetry matrix.
From equality (\ref{TWO_C2_6}), time-independence, energy conservation 
and symplecticity it follows that the polynomial ${\cal F}_{3, m + 1}$
in (\ref{IFB_3}) does not depend on the variable $\sigma$ and is an even 
function of the variables $y$ and $p_y$, and that
the matrix $\,M\,$ in (\ref{IFB_3}) has the following form

\vspace{-0.15cm}
\noindent   
\begin{eqnarray}
M \,=\,
\left(
\begin{array}{cccccc}
r_{11} & r_{12} & 0      & 0      & 0 & r_{16} \\
r_{21} & r_{22} & 0      & 0      & 0 & r_{26} \\
0      & 0      & r_{33} & r_{34} & 0 & 0      \\
0      & 0      & r_{43} & r_{44} & 0 & 0      \\
r_{51} & r_{52} & 0      & 0      & 1 & r_{56} \\
0      & 0      & 0      & 0      & 0 & 1
\end{array}
\right).
\label{ForwMatrix}
\end{eqnarray}

\vspace{-0.15cm}
Using the representation (\ref{IFB_3}) we can state that the map ${\cal M}$
is a $m$-order achromat if, and only if, the matrix of its linear part $M$ 
is dispersion free (i.e. $r_{16}=r_{26}=0$) and the polynomial ${\cal F}_{3, m + 1}$ 
is a function of the variable $\varepsilon$ only.

\vspace{-0.15cm}
\section{DISPERSION DECOMPOSITION}

\vspace{-0.1cm}
Let us consider a midplane symmetric cell with the map 

\vspace{-0.2cm}
\noindent
\begin{eqnarray}
:{\cal M}_c: \,=_3\,
\exp(:{\cal F}_{3,4}^{c}(\mbox{\boldmath $z$}):) :M_c:,
\label{TWO_C_1}
\end{eqnarray}

\vspace{-0.2cm}
\noindent
and let us assume that the cell transfer matrix $M_c$ allows
the solution for the periodic (matched) dispersion to be found.
If we denote by $A$ and $B$ the initial conditions for the periodic
cell dispersion and its derivative respectively,
then the cell transfer matrix can be represented in the form

\vspace{-0.2cm}
\noindent
\begin{eqnarray}
M_c\,=\,D_c\, N_c\, D^{-1}_c ,
\label{TWO_C_3}
\end{eqnarray}

\vspace{-0.2cm}
\noindent
where the matrix $N_c$ is dispersion-free
and its $4 \times 4$ upper left block $\hat{M}_c$
coincides with the corresponding block of
the matrix $M_c$, and the matrix $D_c $ 
can be expressed in the form of a Lie operator as follows

\vspace{-0.2cm}
\noindent   
\begin{eqnarray}
:D_c: \,=\, \exp(:\varepsilon \, (B \, x  \,-\, A \, p_x):).
\label{TWO_C_4}
\end{eqnarray}

Using the dispersion decomposition (\ref{TWO_C_3}) 
the cell transfer map (\ref{TWO_C_1}) can be brought
into the form

\vspace{-0.2cm}
\noindent
\begin{eqnarray}
:{\cal M}_c: \,=_3\,
:D_c:^{-1}
\exp(:{\cal P}_{3,4}^c(\mbox{\boldmath $z$}):) 
\,:N_c:\, :D_c:
\label{TWO_C2_1}
\end{eqnarray}

\vspace{-0.2cm}
\noindent
with 
$\,{\cal P}_{3,4}^c(\mbox{\boldmath $z$})\,=\,
{\cal F}_{3,4}^c(x + A \cdot \varepsilon,\, 
p_x + B \cdot \varepsilon, \,y,\, p_y,\, \varepsilon)$.

Let us now turn our attention to the special features 
of the dispersion decompositions of cells with symmetries. 

The matrix of a system which is mirror symmetric about the $x-y$ plane 
to the original system  is given by

\vspace{-0.2cm}
\noindent
\begin{eqnarray}
M_R \,=\, T_R \, M^{-1} \, T_R,
\label{ReverDef}
\end{eqnarray}

\vspace{-0.2cm}
\noindent
where $T_R = \mbox{diag}(1, -1, 1, -1, -1, 1)$
is the reversion transformation matrix,
and the matrix of the system which is mirror antisymmetric about the $x-y$ plane
to the original system (reversed and then rotated by $180^{\circ}$ around 
the longitudinal axis) is given by

\vspace{-0.2cm}
\noindent
\begin{eqnarray}
M_A \,=\, T_A \, M^{-1} \, T_A, 
\label{CombDef}
\end{eqnarray}

\vspace{-0.2cm}
\noindent
where $T_A = \mbox{diag}(-1, 1, -1, 1, -1, 1)$.

Let us consider a mirror symmetric cell where
the system with the transfer matrix $M$ is followed
by its reversed image and let us assume that the 
$r_{21}$ element of the matrix $M$ is not equal to zero.
Then the cell transfer matrix $M_c = M_R M$
allows dispersion decomposition (\ref{TWO_C_3}) with
$A = -r_{26} \,/\, r_{21}$ and $B = 0$.
The equality $B = 0$ is an remarkable property,
because from it follows the commutation relation 

\vspace{-0.35cm}
\noindent   
\begin{eqnarray}
T_R \, D_c \,=\, D_c \, T_R,
\label{TWO_C_48}
\end{eqnarray}

\vspace{-0.2cm}
\noindent   
which is very important for all our further considerations. 
 
The dispersion decomposition (\ref{TWO_C_3}) is also applicable
to the mirror antisymmetric cell, if one assumes that 
$r_{12} \neq 0$ and takes $A =0$ and
$B = -r_{16} \,/\, r_{12}$.
The commutation relation in this case is the relation
$\,T_A \, D_c \,=\, D_c \, T_A$.

\vspace{-0.2cm}
\section{ABERRATIONS OF PERIODIC SYSTEMS}

\vspace{-0.1cm}
Let us consider a system constructed by a repetition of $n$ identical cells
and let us assume that the cell matrix $M_c$ allows dispersion decomposition.
Then, using representation (\ref{TWO_C2_1}) for the cell map ${\cal M}_c$,
the map of the repetitive $n$-cell system can be expressed as follows

\vspace{-0.2cm}
\noindent
\begin{eqnarray}
:{\cal M}_{nc}: 
\,=\,
:{\cal M}_{c}:^n 
\nonumber
\end{eqnarray}

\vspace{-0.5cm}
\noindent
\begin{eqnarray}
=_3
\exp(:{\cal S}_{3, 4}(D_c^{-1}\mbox{\boldmath $z$})
+ {\cal W}_4(D_c^{-1}\mbox{\boldmath $z$}):) :M_c^n:.
\label{RepSingleLie_3}
\end{eqnarray}

\vspace{-0.2cm}
\noindent
In this representation the aberration functions ${\cal S}_{3,4}$ 
and ${\cal W}_4$ are given by the following formulas

\vspace{-0.25cm}
\noindent
\begin{eqnarray}
{\cal S}_{3, 4}(\mbox{\boldmath $z$}) =
\sum\limits_{m = 0}^{n - 1}
{\cal P}_{3, 4}^c(\hat{M}_c^m \hat{\mbox{\boldmath $z$}},\, \varepsilon),
\label{RepSingleLie_7}
\end{eqnarray}

\vspace{-0.6cm}
\noindent
\begin{eqnarray}
{\cal W}_4(\mbox{\boldmath $z$}) =
\frac{1}{2}
\sum\limits_{l = 0}^{n-2}
\sum\limits_{j = l}^{n - 2}
\big\{
{\cal P}_3^c (\hat{M}_c^l \hat{\mbox{\boldmath $z$}}, \varepsilon),
{\cal P}_3^c (\hat{M}_c^{j + 1} \hat{\mbox{\boldmath $z$}}, \varepsilon)
\big\},
\label{RepSingleLie_8}
\end{eqnarray}

\vspace{-0.2cm}
\noindent
where the binary operation
$\{*,*\}$ is the Poisson bracket.

If, additionally, the basic cell is mirror symmetric with the map ${\cal M}$ 
of its first part given by the Lie factorization

\vspace{-0.25cm}
\noindent
\begin{eqnarray}
:{\cal M}: \,=_3\,
\exp(:{\cal F}_{3,4}(\mbox{\boldmath $z$}):) :M:,
\label{TWO_C_MS_1}
\end{eqnarray}

\vspace{-0.25cm}
\noindent
then the functions ${\cal P}_{3}^c$ and ${\cal P}_{4}^c$ in the formulas
(\ref{RepSingleLie_7}) and (\ref{RepSingleLie_8})
can be re expressed using the commutation relation (\ref{TWO_C_48})
through the function
${\cal P}_{3,4}(\mbox{\boldmath $z$})={\cal F}_{3,4}(D_c\, \mbox{\boldmath $z$})$
as follows

\vspace{-0.2cm}
\noindent
\begin{eqnarray}
{\cal P}_{3}^c(\mbox{\boldmath $z$}) \,= \,
{\cal P}_3(\hat{\mbox{\boldmath $z$}}, \varepsilon)
\,+\,
{\cal P}_3(\hat{T}_R \hat{M}_c \,\hat{\mbox{\boldmath $z$}}, \varepsilon),
\label{TWO_C_MS_2}
\end{eqnarray}

\vspace{-0.4cm}
\noindent
\begin{eqnarray}
{\cal P}_{4}^c(\mbox{\boldmath $z$}) \,= \,
{\cal P}_4(\hat{\mbox{\boldmath $z$}}, \varepsilon)
\,+\,
{\cal P}_4(\hat{T}_R \hat{M}_c \,\hat{\mbox{\boldmath $z$}}, \varepsilon),
\nonumber
\end{eqnarray}

\vspace{-0.4cm}
\noindent
\begin{eqnarray}
+\,
(1\,/\,2)\,
\big\{
{\cal P}_3 (\hat{\mbox{\boldmath $z$}}, \varepsilon),
{\cal P}_3 (\hat{T}_R \hat{M}_c \,\hat{\mbox{\boldmath $z$}}, \varepsilon)
\big\},
\label{TWO_C_MS_3}
\end{eqnarray}

\vspace{-0.2cm}
\noindent
where the $4 \times 4$ matrix $\hat{T}_R$ is the upper left block
of the matrix  $T_R$.

The formulas (\ref{TWO_C_MS_2}) and (\ref{TWO_C_MS_3})
are also applicable to the mirror antisymmetric cell 
with simple exchange of the matrix $\hat{T}_R$ with the matrix $\hat{T}_A$
and thus in the following we will restrict our consideration to
the mirror symmetric case only.

\vspace{-0.25cm}
\section{GROUPS AND ACHROMATS}

\vspace{-0.1cm}
Let us assume that the cell matrix $M_c$ allows dispersion 
decomposition (\ref{TWO_C_3}).
Then, according to the formulas (\ref{RepSingleLie_7}) and (\ref{TWO_C_MS_2}), 
the function ${\cal S}_{3}$, which is responsible
for the second order aberrations of the repetitive $n$-cell system, 
can be represented in the form

\vspace{-0.2cm}
\noindent
\begin{eqnarray}
{\cal S}_{3}(\hat{\mbox{\boldmath $z$}}, \varepsilon) =
n \cdot \mbox{\boldmath ${\cal R}_C$}
\big({\cal P}_{3}^c(\hat{\mbox{\boldmath $z$}}, \varepsilon)\big),
\label{RepSingleLie_10}
\end{eqnarray}

\vspace{-0.4cm}
\noindent
\begin{eqnarray}
\mbox{\boldmath ${\cal R}_C$}\big(f(\hat{\mbox{\boldmath $z$}}, \varepsilon)\big)
= \frac{1}{n}
\sum\limits_{m = 0}^{n - 1}
f(\hat{M}_c^m \hat{\mbox{\boldmath $z$}},\, \varepsilon)
\label{RepSingleLie_9}
\end{eqnarray}

\vspace{-0.2cm}
\noindent
for the basic cell without symmetries and in the form

\vspace{-0.2cm}
\noindent
\begin{eqnarray}
{\cal S}_{3}(\hat{\mbox{\boldmath $z$}}, \varepsilon) =
2 n \cdot
\mbox{\boldmath ${\cal R}_D$}
\big({\cal P}_{3}(\hat{\mbox{\boldmath $z$}}, \varepsilon)\big),
\label{RepSingleLie_20}
\end{eqnarray}

\vspace{-0.4cm}
\noindent
\begin{eqnarray}
\mbox{\boldmath ${\cal R}_D$}(f)
= \frac{1}{2n}
\sum\limits_{j = 0}^{n - 1}
\left(
f(\hat{M}_c^j \hat{\mbox{\boldmath $z$}},\varepsilon)
+
f(\hat{T}_R \hat{M}_c^{j+1} \hat{\mbox{\boldmath $z$}},\varepsilon)
\right)
\label{RepSingleLie_19}
\end{eqnarray}

\vspace{-0.2cm}
\noindent
for the mirror symmetric basic cell.

\vspace{-0.1cm}
\subsection{Appearance of the Cyclic Group $C_n$}

For the general midplane symmetric cell the polynomial ${\cal P}_{3}^c$ can have 
as much as 18 nonzero monomials responsible for the independent second order
transverse aberrations. Why should one expect that the polynomial ${\cal S}_3$ 
defined by the equality (\ref{RepSingleLie_10}) has a smaller number of them, i.e. 
why should one expect that the map of the $n$-cell system has less independent 
second order aberrations than the cell map? No reason is seen for that in the 
case of an arbitrary matrix $\hat{M}_c$. The situation will change if the matrices

\vspace{-0.2cm}
\noindent
\begin{eqnarray}
I,\,\,\hat{M}_c,\,\ldots,\,\, \hat{M}_c^{n-1}
\label{CYC_GR_111}
\end{eqnarray}

\vspace{-0.2cm}
\noindent
will form a finite matrix group. 
Then the operator 
$\mbox{\boldmath ${\cal R}_C$}$ becomes the Reynolds (averaging) operator of
this group and, therefore, the polynomial ${\cal S}_3$ is not an arbitrary
polynomial anymore. It becomes (as a result of averaging) a polynomial which 
is invariant under the group action.
Moreover, the functions ${\cal S}_4$ and ${\cal W}_4$, which are responsible 
for the third order aberrations, will satisfy the identities

\vspace{-0.2cm}
\noindent
\begin{eqnarray}
{\cal S}_{4}(\hat{\mbox{\boldmath $z$}}, \varepsilon) =
\mbox{\boldmath ${\cal R}_C$}
\big({\cal S}_{4}(\hat{\mbox{\boldmath $z$}}, \varepsilon)\big),
\label{RepSingleLie_101}
\end{eqnarray}

\vspace{-0.4cm}
\noindent
\begin{eqnarray}
{\cal W}_4(\hat{\mbox{\boldmath $z$}}, \, \varepsilon) =
\mbox{\boldmath ${\cal R}_C$}
\big({\cal W}_4(\hat{\mbox{\boldmath $z$}}, \varepsilon)\big)
\nonumber
\end{eqnarray}

\vspace{-0.7cm}
\noindent
\begin{eqnarray}
+
\sum\limits_{m = 0}^{n - 2}
(n - m - 1)
\big\{
{\cal P}_3^c (\hat{M}_c^m \hat{\mbox{\boldmath $z$}}, \varepsilon),
\,
\mbox{\boldmath ${\cal R}_C$}
\big({\cal P}_3^c(\hat{\mbox{\boldmath $z$}}, \varepsilon)\big)
\big\},
\label{RepThirdOrder_3}
\end{eqnarray}

\vspace{-0.2cm}
\noindent
from which it follows that if the  function ${\cal S}_{3}$
does not depend on the variables $\hat{\mbox{\boldmath $z$}}$, 
then the functions ${\cal S}_4$ and ${\cal W}_4$
are also invariants under the group action.

It is clear that the only group which the matrices (\ref{CYC_GR_111})
can form is a cyclic group, and, in order to have a cyclic group
of order $n$ and not several copies of some cyclic group of
smaller dimension, we assume in the following that

\vspace{-0.15cm}
\noindent
\begin{eqnarray}
\hat{M}_c^n = I
\;\;\;
\mbox{and}
\;\;\;
\hat{M}_c^m \neq I
\;\;\;
\mbox{for}
\;\;\;
m = 1, \ldots, n-1.
\label{CYC_GR_1}
\end{eqnarray}

\vspace{-0.2cm}
\noindent
Besides that we assume that the $2 \times 2$ horizontal
focusing block of the matrix $\hat{M}_c$ is not equal to the
identity matrix, which, when combined with the equality $\hat{M}_c^n = I$,
gives sufficient conditions for the existence of the periodic cell 
dispersion and for the property of the overall transfer matrix of 
the $n$-cell system to be a linear achromat.

\vspace{-0.1cm}
\subsection{Appearance of the Dihedral Group $D_n$}

If the basic cell is mirror symmetric, then 
good cancellation of aberrations can be achieved if 
the group will be formed by the set of matrices

\vspace{-0.2cm}
\noindent
\begin{eqnarray}
I,\,\,\hat{M}_c,\,\ldots,\,\, \hat{M}_c^{n-1},\,\,
\hat{T}_R \hat{M}_c, \,\,\ldots, \,\,\hat{T}_R \hat{M}_c^n.
\label{CYC_GR_222}
\end{eqnarray}

\vspace{-0.2cm}
\noindent
What kind of group could it be?
One can check that while the last $n$ matrices in (\ref{CYC_GR_222})
are antisymplectic, the first $n$ matrices coincide
with the set (\ref{CYC_GR_111}) and are symplectic, and thus
must form a subgroup. So let us assume again that the conditions
(\ref{CYC_GR_1}) are satisfied and the subset 
of the first $n$ matrices of the set (\ref{CYC_GR_222}) 
is isomorphic to the cyclic group $C_n$.
Then the set (\ref{CYC_GR_222}) becomes isomorphic
to the dihedral group $D_n$ consisting of $n$ reflections,
$n-1$ rotations and the identity transformation, and the analogies of 
the formulas (\ref{RepSingleLie_101}) and (\ref{RepThirdOrder_3}) 
take on the forms

\vspace{-0.1cm}
\noindent
\begin{eqnarray}
{\cal S}_4(\hat{\mbox{\boldmath $z$}}, \, \varepsilon) =
\mbox{\boldmath ${\cal R}_D$}\big(
{\cal S}_4(\hat{\mbox{\boldmath $z$}}, \varepsilon)
\big),
\label{DihGroup_N_1}
\end{eqnarray}

\vspace{-0.3cm}
\noindent
\begin{eqnarray}
{\cal W}_4(\hat{\mbox{\boldmath $z$}}, \, \varepsilon) =
\mbox{\boldmath ${\cal R}_D$}\big(
{\cal W}_4(\hat{\mbox{\boldmath $z$}}, \varepsilon)
\big)
\nonumber
\end{eqnarray}

\vspace{-0.7cm}
\noindent
\begin{eqnarray}
+2 \sum\limits_{j = 0}^{n - 2}
(n - j - 1)
\big\{{\cal P}_3^c (\hat{M}_c^j \hat{\mbox{\boldmath $z$}}, \varepsilon),
\mbox{\boldmath ${\cal R}_D$}
\big({\cal P}_3(\hat{\mbox{\boldmath $z$}}, \varepsilon)\big)
\big\}.
\label{DihGroup_N_10}
\end{eqnarray}

\vspace{-0.45cm}
\subsection{Optimal Choice of the Cell Phase Advances}

From the equalities (\ref{RepSingleLie_10}), (\ref{RepSingleLie_101})
and (\ref{RepThirdOrder_3}) it follows that the number of constraints 
which one has to satisfy (typically, with usage of multipoles) in order 
to make repetitive $n$-cell system a second or third order achromat is 
defined by the number of distinct invariant homogeneous polynomials in
the variables $\hat{\mbox{\boldmath $z$}}$  
of the cyclic group $C_n$ generated by the matrix $\hat{M}_c$.
As an abstract object the group $C_n$ is unique and
for all possible matrices $\hat{M}_c$ 
satisfying (\ref{CYC_GR_1}) we have groups which are isomorphic to each other,
but not all of them are conjugate. Thus as groups of symmetries they can be
distinct and can have different number of 
uncanceled second ($a_2$) and third ($a_3$) order aberrations,
and this depends on the choice of the periodic 
cell phase advances $\mu_{x,y}^c$.
Due to our assumptions about the matrix $\hat{M}_c$ 

\vspace{-0.22cm}
\noindent
\begin{eqnarray}
\mu_{x, y}^c \,=\, 2 \pi \,q_{x,y} \,/\, n
\;\;\;(\mbox{mod} \, 2\pi)
\label{RepCellTunes_1_1}
\end{eqnarray}

\vspace{-0.22cm}
\noindent
for some $q_x = 1,\ldots,n-1$ and some $q_y = 0,\ldots,n-1$ such
that the numbers $n,\,q_x,\,q_y$
have no common positive factor other than 1.

For arbitrary $\mu_{x, y}^c$ the numbers $a_2$ and $a_3$ cannot
be smaller than two and five respectively (chromatic and nonlinear 
contributions to the cell tunes which cannot be canceled only by group averaging), 
and their actual values depend on
the number of resonances which $\mu_{x, y}^c$ satisfy and
which are not forbidden by the midplane symmetry.
The phase advances which are optimal for the
construction of the repetitive third order achromats are given in 
the table 1 for $n=2,\ldots,10$. Note that in the last column of 
this table only the 
generating pairs $(q_x,\,q_y)$ are shown. In order to obtain all 
optimal pairs one has to multiply generating pairs by all positive integers
which are smaller than $n$ and are coprime to $n$ using
modulo $n$ arithmetic.
For example, for $n=8$ we have in this table the pair 
$(q_x,\,q_y) = (1, 3)$.
This pair, when multiplied by $3$, $5$ and $7$ 
using modulo 8 arithmetic,
generates three more optimal pairs $(3, 1)$, $(5, 7)$ and $(7, 5)$.

\begin{table}[t]
\vspace{-0.3cm}
\caption{
\label{tab:table1}
The pairs $(q_x, q_y)$ which minimize the sum $a_2+a_3$ of the number of 
independent transverse second and third order aberrations
of the midplane symmetric $n$-cell system.\vspace{-0.3cm}}
\begin{center}
\begin{tabular}{cccc}
\hline
\hline
 $n$ & $a_2$ & $a_3$ & $(q_x,\,q_y)$\\
\hline
2  & 6 & 25 & (1, 0), (1, 1)\\
3  & 6 & 11 & (1, 1), (1, 2)\\
4  & 2 & 13 & (1, 1), (1, 3)\\
5  & 2 &  7 & (1, 1), (1, 4)\\
6  & 2 &  7 & (1, 1), (1, 2), (1, 4), (1, 5)\\
7  & 2 &  5 & (1, 2), (1, 5)\\
8  & 2 &  7 & (1, 1), (1, 2), (1, 3),(1, 5), (1, 6), (1, 7)\\
9  & 2 &  5 & (1, 2), (1, 3), (1, 6), (1, 7)\\
10 & 2 &  5 & (1, 2), (1, 3), (1, 7), (1, 8)\\
\hline
\hline
\end{tabular}
\end{center}
\vspace{-0.7cm}
\end{table}

If the basic cell is mirror symmetric, then,
for the same cell tunes, the numbers of 
remaining transverse second ($\hat{a}_2$) and third ($\hat{a}_3$) 
order aberrations are given by the relations

\vspace{-0.15cm}
\noindent
\begin{eqnarray}
\hat{a}_2 = 2 + (a_2 - 2) \,/\, 2,
\;\;\;\;
\hat{a}_3 = 5 + (a_3 - 5) \,/\, 2.
\label{AbbMirS}
\end{eqnarray}

\vspace{-0.15cm}
\noindent
One sees that if $a_2 > 2$ or $a_3 > 5$, then 
the usage of the mirror symmetry reduces 
the number of aberrations left for correction by
multipoles. So if the number of the independent multipole 
families required for the aberration correction is of concern, 
then it could be better to use achromats based on the $D_n$ group.
But if the total number of multipole magnets has to be minimized,
then achromats utilizing the $C_n$ group perform better,
because in the mirror symmetric case one has to put multipoles
into the basic cell in such a way that the symmetry is preserved.

\end{document}